\documentclass[12pt]{article}
\usepackage{epsfig}
\begin{document}
\begin{center}
{\Large {\bf Standard Model with the additional $Z_6$ symmetry on the lattice}}

\vskip-40mm \rightline{\small ITEP-LAT/2005-04} \vskip 30mm

{
\vspace{1cm}
{B.L.G.~Bakker$^a$, A.I.~Veselov$^b$, M.A.~Zubkov$^b$  }\\
\vspace{.5cm} { \it $^a$ Department of Physics and Astronomy, Vrije
Universiteit, Amsterdam,
The Netherlands \\
$^b$ ITEP, B.Cheremushkinskaya 25, Moscow, 117259, Russia }}
\end{center}

\begin{abstract}
An additional $Z_6$ symmetry hidden in the fermion and Higgs sectors of the
Standard Model  has been found recently\cite{BVZ2003}. A lattice regularization
of the Standard Model was constructed that possesses this symmetry. In
\cite{BVZ2004} we have reported our results on the numerical simulation of the
electroweak sector of the model. In this paper we report our results on the
numerical simulation of the full ($SU(3)\otimes SU(2) \otimes U(1)$ ) model.
The  phase diagram of the model has been investigated using static quark and
lepton potentials. Various types of monopoles have been constructed. Their
densities appear to be sensitive to the phase transition lines. Differences
between the realizations of the Standard Model which do or do not possess
the mentioned $Z_6$ symmetry, are discussed.
\end{abstract}


\section{Introduction}
\label{sect.1}

Until recently it was thought that all the symmetries of the Standard
Model (SM), which must be used when dealing with its discretization,
are known.  However, in \cite{BVZ2003} it was shown that there exists
an additional $Z_6 = Z_2 \otimes Z_3$ symmetry in the fermion and Higgs
sectors of the SM. It  is connected to the centers $Z_3$ and $Z_2$ of
the $SU(3)$ and $SU(2)$ subgroups\footnote{The emergence of $Z_6$
symmetry in the SM and its supersymmetric extension was
independently considered in a different context in \cite{Z6}.}. The gauge
sector of the SM (in its discretized form) was redefined in such a way
that it has the same naive continuum limit as the original one, while
keeping the mentioned symmetry.  The resulting model differs from the
conventional SM via its symmetry properties. Therefore we expect, that
nonperturbatively these two models may represent different physics.

Investigation of the electroweak sector of the SM with the additional
$Z_6$ symmetry shows, that there are indeed certain differences between
this discretization and the conventional one \cite{BVZ2004}. Namely, it
has been found that the phase transition lines corresponding to the
$U(1)$ and $SU(2)$ degrees of freedom join in a triple point, forming a
common line. In contrast to this, in the conventional model the phase
transition line corresponding to $SU(2)$ degrees of freedom has an
endpoint and the transition becomes continuous in a certain region of
coupling constants \cite{SU2U1}. In this paper we report our results on
the full SM (including $SU(3)$ degrees of freedom) and claim that the
same phenomenon takes place here. Now the $SU(3)$, $SU(2)$ and $U(1)$
degrees of freedom are connected via their centers. This, in our
opinion, is the reason why the phase transition lines corresponding to
the phase transitions in pure $U(1)$ and $SU(2)$ models again join
together forming a common line. It turns out that $SU(3)$ fields
experience this common phase transition as well.

This paper is organized as follows. In the next section we summarize the
formulation of the SM in terms of link variables and demonstrate
the emergence of an additional $Z_6$ symmetry in its fermion and Higgs sectors.
In Sect.~\ref{sect.3} we detail the model with explicit $Z_6$ symmetry on the
lattice, while in Sect.~\ref{sect.4} we recall the definition of the maximal
center projection. The next section contains the definitions of the quantities
we measure on the lattice; it is followed by Sect.~\ref{sect.6} where we show
our numerical results. We end with a summary.

\section{$Z_6$ symmetry in the Standard Model}
\label{sec.2}

In this section we remind the reader of what we call the additional
$Z_6$ symmetry. The SM contains the following variables:

1. The gauge field ${\cal U} = (\Gamma, U, \theta)$, where
\begin{eqnarray}
 \Gamma \in SU(3), \quad U \in SU(2), \quad e^{i\theta} \in U(1),
\end{eqnarray}
realized as link variables on the lattice.

2. A scalar doublet
\begin{equation}
 \Phi^{\alpha}, \;\alpha = 1,2.
\end{equation}

3. Anticommuting spinor variables, representing leptons and quarks:
\begin{equation}
 \left(
 \begin{array}{ccc}
  \nu_e & \nu_{\mu} &  \nu_{\tau}\\
  e & \mu & \tau ,
 \end{array}
 \right) , \quad
 \left(
 \begin{array}{ccc}
 u & c & t \\
 d & s & b
 \end{array}
 \right) .
\end{equation}

The action has the form
\begin{equation}
 S = S_g + S_H + S_f,
\end{equation}
where we denote the fermion part of the action by $S_{f}$, the pure
gauge part is denoted by $S_g$, and the scalar part of the action by
$S_H$.

In {\it any} lattice realization of $S_H$ and $S_f$ both these terms
depend upon link variables $\cal U$ considered in the representations
corresponding to quarks, leptons, and the Higgs scalar field,
respectively. Therefore $\cal U$ appears in the combinations shown in
the table.
\begin{table}
\label{tab.01}
\begin{center}
\begin{tabular}{|c|l|}
\hline
$U\, e^{-i\theta}$ & {\rm left-handed leptons} \\
\hline
$e^{-2 i \theta}$ & {\rm right-handed leptons} \\
\hline
$ \Gamma \, U \, e^{ \frac{i}{3} \theta}$ & {\rm left-handed quarks}\\
\hline
$ \Gamma \, e^{ -\frac{2i}{3} \theta}$ &{\rm right-handed $d$, $s$, and, $b$ - quarks} \\
\hline
$ \Gamma \, e^{ \frac{4i}{3} \theta}$ &{\rm right-handed $u$, $c$, and, $t$ - quarks} \\
\hline
$  U \, e^{  i \theta}$ &{\rm the Higgs scalar field}\\
\hline
\end{tabular}
\end{center}
\end{table}
Our observation is that {\it all} the listed combinations are invariant
under the following transformations:
\begin{eqnarray}
 U & \rightarrow & U e^{-i\pi N}, \nonumber\\
 \theta & \rightarrow & \theta +  \pi N, \nonumber\\
 \Gamma & \rightarrow & \Gamma e^{(2\pi i/3)N},
\label{symlat}
\end{eqnarray}
where $N$ is an arbitrary integer link variable. It represents a
three-dimensional hypersurface on the dual lattice. Both $S_H$ and
$S_f$ (in {\it any} realization) are invariant under the simultaneous
transformations (\ref{symlat}). This symmetry reveals the
correspondence between the centers of the $SU(2)$ and $SU(3)$ subgroups
of the gauge group.

After integrating out fermion and scalar degrees of freedom any
physical variable should depend upon gauge invariant quantities only.
Those are the Wilson loops: $\omega_{SU(3)}({\cal C}) = {\rm Tr}
\Pi_{{\rm link} \in {\cal C}} \Gamma_{\rm link}$, $\omega_{SU(2)}({\cal
C}) = {\rm Tr} \Pi_{{\rm link} \in {\cal C}} U_{\rm link}$, and
$\omega_{U(1)}({\cal C}) = \Pi_{{\rm link} \in {\cal C}} {\rm
exp}(\frac{i}{3} \theta_{\rm link})$. Here $\cal C$ is an arbitrary
closed contour on the lattice (with self - intersections allowed).
These Wilson loops are trivially invariant under the transformation
(\ref{symlat}) with the field $N$ representing a {\it closed}
three-dimensional hypersurface on the dual lattice. Therefore the
nontrivial part of the symmetry (\ref{symlat}) corresponds to a closed
two-dimensional surface on the dual lattice that is the boundary of the
hypersurface represented by $N$. Then in terms of the gauge invariant
quantities $\omega$ the transformation (\ref{symlat}) acquires the
form:
\begin{eqnarray}
 \omega_{U(1)}({\cal C}) & \rightarrow &
 {\rm exp}(-i \mbox{\small $\frac{1}{3}$} \pi {\bf L} ({\cal C}, \Sigma))
 \, \omega_{U(1)}({\cal C}) \nonumber\\
 \omega_{SU(2)}({\cal C}) &
 \rightarrow & {\rm exp}( i \pi {\bf L} ({\cal C}, \Sigma))
 \, \omega_{SU(2)}({\cal C}) \nonumber\\
 \omega_{SU(3)}({\cal C}) & \rightarrow &
 {\rm exp}( i \mbox{\small $\frac{2}{3}$} \pi {\bf L}
 ({\cal C}, \Sigma))\, \omega_{SU(3)}({\cal C}) \label{sym}
\end{eqnarray}
Here $\Sigma$ is an arbitrary closed surface (on the dual lattice) and
${\bf L} ({\cal C}, \Sigma)$ is the integer linking number of this
surface and the closed contour $\cal C$. From (\ref{sym}) it  follows,
that the symmetry is of $Z_6$ type.

\section{The model under investigation}
\label{sect.3}

It is obvious that the pure gauge-field part of the action in its
conventional continuum formulation (or, say, in lattice Wilson
formulation) is not invariant under (\ref{sym}). However, the lattice
realization of the pure gauge field term of the action can be
constructed in such a way that it also preserves the mentioned
symmetry. For the reasons listed in \cite{BVZ2003} we consider it in
the following form:
\begin{eqnarray}
 S_g & = & \beta \sum_{\rm plaquettes}
 \{2(1-\mbox{${\small \frac{1}{2}}$} {\rm Tr}\, U_p \cos \theta_p)+
\nonumber \\
 && +\;(1-\cos 2\theta_p) \nonumber \\
 && +\;6[1-\mbox{${\small \frac{1}{6}}$} {\rm Re Tr}
 \,\Gamma_p {\rm Tr}\, U_p {\rm exp} (i\theta_p/3)]
\nonumber\\
 && +\;3[1-\mbox{${\small \frac{1}{3}}$} {\rm Re Tr}
 \, \Gamma_p {\rm exp} (-2i\theta_p/3)]
\nonumber \\
 && +\;3[1-\mbox{${\small \frac{1}{3}}$} {\rm Re Tr}
 \, \Gamma_p {\rm exp} (4i\theta_p/3)]\},
\label{Act}
\end{eqnarray}
where the sum runs over the elementary plaquettes of the lattice. Each
term of the action Eq.~(\ref{Act}) corresponds to a parallel
transporter along the boundary  $\partial p$ of plaquette $p$. The
corresponding plaquette variables constructed of lattice gauge fields
are $U_p = \omega_{SU(2)}(\partial p)\, , \Gamma_p =
\omega_{SU(3)}(\partial p)\, $, and $ \theta_p = {\rm Arg}\,
\omega_{U(1)}(\partial p)$.

The potential for the scalar field is considered in its simplest form
\cite{BVZ2004} in the London limit, i.e., in the limit of infinite bare
Higgs mass. After fixing the unitary gauge we obtain:
\begin{eqnarray}
 S_H & = & \gamma \sum_{xy}[1 - Re(U^{11}_{xy} e^{-i\theta_{xy}})].
\end{eqnarray}
The following variables are (naively) considered as creating a photon, $Z$
boson, and $W$ boson respectively:
\begin{eqnarray}
 A_{xy} & = & A^{\mu}_{x} \; = \,[{\rm Arg} U_{xy}^{11} + \theta_{xy}]
 \,{\rm mod} \,2\pi, \nonumber\\
 Z_{xy} & = & Z^{\mu}_{x} \; = \,[{\rm Arg} U_{xy}^{11} - \theta_{xy}]
 \,{\rm mod} \,2\pi, \nonumber\\
 W_{xy} & = & W^{\mu}_{x} \,= \,U_{xy}^{12} e^{i\theta_{xy}}.
\end{eqnarray}
Here, $\mu$ represents the direction $(xy)$. After fixing the unitary
gauge the electromagnetic $U(1)$ symmetry remains:
\begin{eqnarray}
 U_{xy} & \rightarrow & g^\dag_x U_{xy} g_y, \nonumber\\
 \theta_{xy} & \rightarrow & \theta_{xy} +  \alpha_y/2 - \alpha_x/2,
\end{eqnarray}
where $g_x = {\rm diag} (e^{i\alpha_x/2},e^{-i\alpha_x/2})$. The fields $A$,
$Z$, and $W$ transform as follows:
\begin{eqnarray}
 A_{xy} & \rightarrow & A_{xy} + \alpha_y - \alpha_x, \nonumber\\
 Z_{xy} & \rightarrow & Z_{xy}, \nonumber\\
 W_{xy} & \rightarrow & W_{xy}e^{-i\alpha_x}.
\label{T}
\end{eqnarray}

We consider our model in quenched approximation, i.e., we neglect the
effect of virtual fermion loops. Therefore the particular form of $S_f$
is not of interest for us at this stage.

In order to extract physical information from the $SU(3)$ fields in a
particulary simple way we use the so-called indirect Maximal Center
Projection (see, for example, \cite{Greensite,BVZ1999}).

\section{The Maximal Center Projection.}
\label{sect.4}

The Maximal Center Projection makes the link matrix $\Gamma$ as close
as possible to the elements of the center $Z_3$ of $SU(3)$: $ Z_3 =
\{{\rm diag}(\mathrm{e}^{(2\pi i /3) N}, \mathrm{e}^{(2\pi i /3) N},
\mathrm{e}^{(2\pi i /3) N}\}$, where $N \in \{1, 0, -1\}$.  The
procedure works as follows.

First, make the functional
\begin{equation}
Q_1 = \sum_{\mathrm{links}} (|\Gamma_{11}| + |\Gamma_{22}| + |\Gamma_{33}|)
\end{equation}
maximal with respect to the gauge transformations $\Gamma_{xy}
\rightarrow g^\dag_x \Gamma_{xy} g_y$, thus fixing the Maximal Abelian
gauge. As a consequence every link matrix becomes almost diagonal.

Secondly, to make this matrix as close as possible to the center of
$SU(3)$, make the phases of the diagonal elements of this matrix
maximally close to each other. This is done by minimizing the
functional
\begin{eqnarray}
 Q_2 & = &
 \sum_{\mathrm{links}}\{[1-\cos({\rm Arg}(\Gamma_{11})-{\rm Arg}(\Gamma_{22}))]
     + [1-\cos({\rm Arg}(\Gamma_{11})-{\rm Arg}(\Gamma_{33}))]
\nonumber \\
 &&  +\; [1-\cos({\rm Arg}(\Gamma_{22})-{\rm Arg}(\Gamma_{33}))]\}.
\end{eqnarray}
with respect to the gauge transformations. This gauge condition is
invariant under the central subgroup $Z_3$ of $SU(3)$.

In our model $SU(3)$ fields are connected with the $U(1)$ and $SU(2)$
fields via the center of the gauge group. Therefore, instead of the center
vortices and center monopoles we define various kinds of monopole -
like fields. The definitions of these fields includes the following
integer-valued link variable $N$ (defined after fixing the Maximal
Center gauge):
\begin{eqnarray}
 N_{xy}=0 &{\rm if}& ({\rm Arg}(\Gamma_{11})+{\rm Arg}(\Gamma_{22})+{\rm Arg}
 (\Gamma_{33}))/3 \in \; ]-\pi/3, \pi/3],
\nonumber \\
 N_{xy}=1 &{\rm if}& ({\rm Arg}(\Gamma_{11})+{\rm Arg}(\Gamma_{22})+{\rm Arg}
 (\Gamma_{33}))/3 \in \; ]\pi/3, \pi],
\nonumber \\
 N_{xy}=-1 &{\rm  if}& ({\rm Arg}(\Gamma_{11})+{\rm Arg}(\Gamma_{22})+{\rm Arg}
 (\Gamma_{33}))/3 \in \; ]-\pi, -\pi/3].
\end{eqnarray}
In other words, $N=0$ if $\Gamma$ is close to $1$,  $N=1$ if $\Gamma$ is
close to \noindent $\mathrm{e}^{2\pi i/3}$ and  $N=-1$ if $\Gamma$ is
close to $\mathrm{e}^{-2\pi i/3}$.

Next, we define the following link fields
\begin{eqnarray}
 C^1_{xy} & = & \,\left[\frac{2\pi}{3}N_{xy} +
 {\rm Arg} U_{xy}^{11} + \frac{1}{3}\theta_{xy}\right] \,{\rm mod} \,2\pi,
\nonumber\\
 C^2_{xy} & = &  \,\left[\frac{2\pi}{3}N_{xy} - \frac{2}{3}\theta_{xy}\right]
 \,{\rm mod} \,2\pi,
\nonumber\\
 C^2_{xy} & = & \,\left[\frac{2\pi}{3}N_{xy} + \frac{4}{3}\theta_{xy}\right]
 \,{\rm mod} \,2\pi.
\end{eqnarray}
These fields correspond to the last three terms of Eq.~(\ref{Act}). Their
construction comes from the representation of $\Gamma$ as
a product of $\exp((2\pi i / 3) N)$  and $V$, where  $V$ is the
$SU(3)/Z_3$ variable $({\rm Arg}(V_{11})+{\rm Arg}(V_{22})+{\rm
Arg}(V_{33}))/3 \in \; ]-\pi/3, \pi/3]$.  Thus $\Gamma=\exp((2\pi i/ 3)
N) V$. We expect, that (\ref{Act}) suppresses $V_{\rm plaq}$ and
$C^i_{\rm plaq}, i = 1,2,3$ while the fields $N$, $\frac{\theta}{3}$,
and $U_{11}$ (being considered independently of each other) are expected
to be disordered. This assumption is justified by the numerical
simulations.

\section{Quantities to be measured}
\label{sect.5}

We investigated five types of monopoles. The monopoles, which carry
information about colored fields are extracted from $C^i$:
\begin{equation}
 j_{C^i} = \frac{1}{2\pi} {}^*d([d C^i]{\rm mod}2\pi) .
\end{equation}
(Here we used the notations of differential forms on the lattice. For
a definition of those notations see, for example, ~\cite{forms}.)

Pure $U(1)$ monopoles, corresponding to the second term in (\ref{Act}), are
extracted from $2\theta$:
\begin{equation}
 j_{2\theta} = \frac{1}{2\pi} {}^*d([d 2\theta]{\rm mod}2\pi).
\end{equation}
The electromagnetic monopoles, corresponding to the first term in (\ref{Act}),
are:
\begin{equation}
 j_{A} = \frac{1}{2\pi} {}^*d([d A]{\rm mod}2\pi) .
\end{equation}

The density of the monopoles is defined as follows:
\begin{equation}
 \rho = \left\langle \frac{\sum_{\rm links}|j_{\rm link}|}{4L^4} \right\rangle,
\end{equation}
where $L$ is the lattice size.
To understand the dynamics of external charged particles, we consider the
Wilson loops defined in the fermion representations listed above (in the
table):
\begin{eqnarray}
 {\cal W}^{\rm L}_{\rm lept}(l) & = &
 \langle {\rm Re} {\rm Tr} \,\Pi_{(xy) \in l} U_{xy} e^{-i\theta_{xy}}\rangle,
\nonumber\\
 {\cal W}^{\rm R}_{\rm lept}(l) & = &
 \langle {\rm Re} \Pi_{(xy) \in l} \, e^{-2i\theta_{xy}}\rangle ,
\nonumber\\
 {\cal W}^{\rm L}_{ {\rm quarks}}(l) & = & \langle {\rm Re} \Pi_{(xy) \in l}
 \, \Gamma_{xy} \, U_{xy}\, e^{\frac{i}{3}\theta_{xy}}\rangle ,
\nonumber\\
 {\cal W}^{\rm R}_{{\rm down} \, {\rm quarks}}(l) & = &
 \langle {\rm Re} \Pi_{(xy) \in l} \, \Gamma_{xy} \,
 e^{-\frac{2i}{3}\theta_{xy}}\rangle ,
\nonumber\\
 {\cal W}^{\rm R}_{{\rm up} \, {\rm quarks}}(l) & = &
 \langle {\rm Re} \Pi_{(xy) \in l} \, \Gamma_{xy} \,
 e^{\frac{4i}{3}\theta_{xy}}\rangle .
\label{WL}
\end{eqnarray}
Here $l$ denotes a closed contour on the lattice. We consider the following
quantity constructed from the rectangular Wilson loop of size $a\times t$:
\begin{equation}
 {\cal V}(a) = \lim_{t \rightarrow \infty}
 \frac{  {\cal W}(a\times t)}{{\cal W}(a\times (t+1))}.
\end{equation}
A linear behavior of ${\cal V}(a)$ would indicate the existence of a
charge - anti charge string with nonzero tension.

\section{Numerical results}
\label{sect.6}

In our calculations we investigated lattices $L^4$ for $L = 6$, $L =
12$, and $L = 16$ with symmetric boundary conditions.

\begin{figure}
\begin{center}
\epsfig{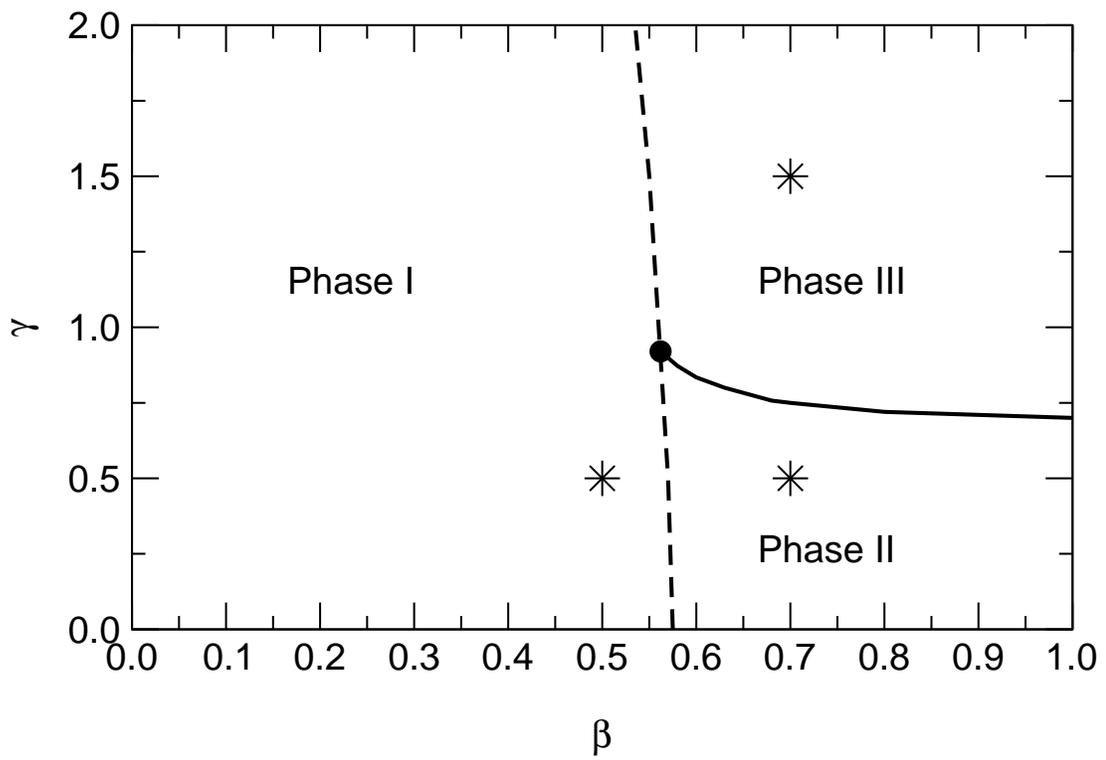}
\caption{\label{fig.1}The phase diagram of the model in the
 $(\beta, \gamma)$-plane.}
\end{center}
\end{figure}

We summarize our qualitative results in the phase diagram represented
in Fig.~\ref{fig.1}. The model contains three phases. The first one (I)
is a phase, in which the dynamics of external leptons is confinement -
like, i.e. is similar to that of external charges in QCD with dynamical
fermions. In the second phase (II) the behavior of left-handed leptons
is confinement-like, while for right-handed ones it is not. The last
one (III) is the Higgs phase, in which no confining forces between
leptons are observed at all. In all three phases there is the
confinement of all external quark fields (left quarks, right up quarks,
right down quarks).

\begin{figure}
 \epsfig{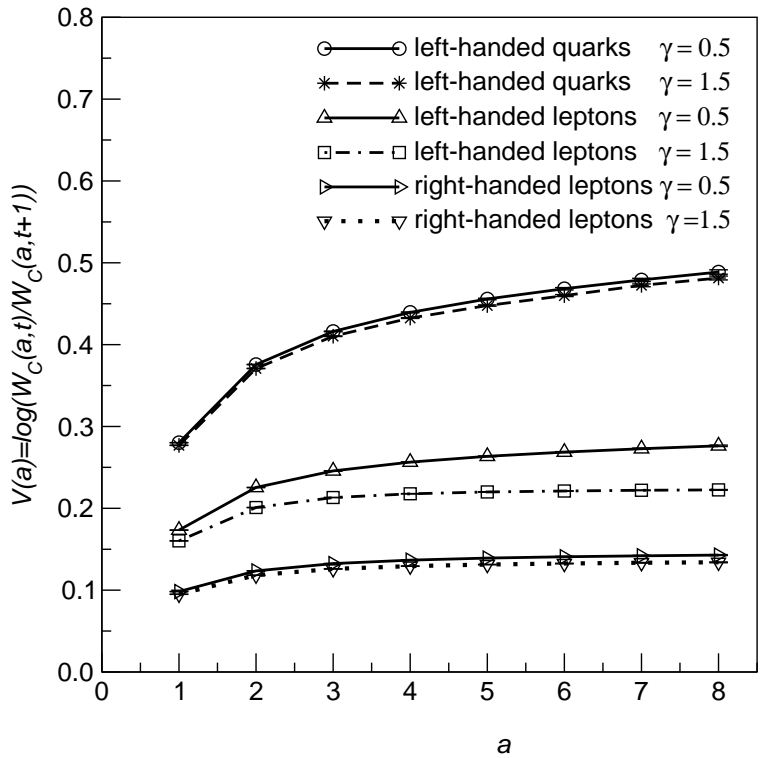}
 \caption{\label{fig.2} ${\cal V}_L(a)$ calculated
  at $\beta = 0.7$. Here the potentials are extracted from
  $ {\cal W}^{\rm L}_{ {\rm quarks}}$(left - handed quarks), $ {\cal W}^{\rm L}_{ {\rm lept}}$ (left - handed  leptons),
and $ {\cal W}^{\rm R}_{ {\rm lept}}$ (right - handed leptons).}
\end{figure}

This is illustrated by Figs.~\ref{fig.2}, in which we show ${\cal V}(a)$
extracted from the Wilson loops Eq.~(\ref{WL}) at two typical points that
belong to phases II ($\gamma = 0.5$) and III  ($\gamma = 1.5$) of the model
(the behavior of all potentials in the phase I is confinement - like). We
represent here the potential for only one colored Wilson loop, i.e. for $ {\cal
W}^{\rm L}_{ {\rm quarks}}$, because the string tension extracted from the
other two potentials coincides with the string tension extracted from the
potential represented in the figure within the errors. This is, of course,
exactly what we have expected: string tensions for different types of quarks
are equal to each other. Thus, the potential, extracted from the colored
fields, possesses linear behavior in all phases, indicating appearance of
confinement of quarks.

By making a linear fit to the lepton potentials at values $a \ge 5$ we
found that only in the case of left-handed leptons the value of the
string tension is much larger than its statistical uncertainty in phase
II. For left-handed leptons in the Higgs phase and right-handed leptons
in both phases, the uncertainty in the values of the string tension
turns out to be larger than about 24\% of its value. In these
cases we do not consider the string tension to be significantly
different from zero. However, as for QCD with dynamical fermions or the
$SU(2)$ fundamental Higgs model \cite{Montvay,EW_T}, these results do
not mean that confinement of leptons occurs. The charge - anti charge
string must be torn by virtual charged scalar particles, which are
present in the vacuum due to the Higgs field. Thus ${\cal V}(a)$ may be
linear only at sufficiently small distances, while starting from some
distance it must not increase, indicating the breaking of the string.
Unfortunately the accuracy of our measurements does not allow us to
observe this phenomenon in detail.

\begin{figure}
 \epsfig{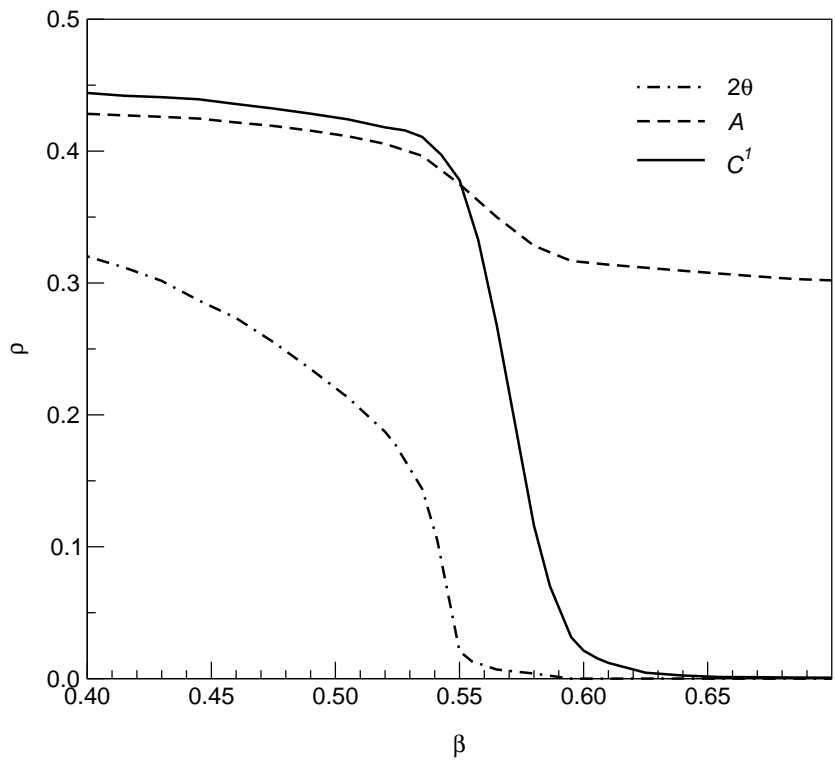}
 \caption{\label{fig.3} The monopole densities (constructed of the link fields
 $A, 2\theta$, and $C^1$)
versus $\beta$ at fixed $\gamma = 0.5$.}
\end{figure}

The connection between the properties of monopoles and the phase
structure of the model is illustrated by Figs.~\ref{fig.3}, which shows
the monopole density versus $\beta$ at fixed $\gamma = 0.5$. Again we
represent here only one type of the three monopoles, which have colored
origin. Namely, we consider $j_{C^1}$. (Behavior of the others is
similar.) One can see, that the density of the $2\theta$ - monopoles as
well as $C^1$ - monopoles falls sharply in phase II, while the
electromagnetic monopole density does not.

We note here, that according to our measurements the electromagnetic
monopole density falls to zero while shifting from phases I to III. The
colored monopoles and $2\theta$ - monopole densities fall sharply in
the phase III as well. Thus monopoles composed of colored fields feel
the phase transition, which are due, according to our intuition, to the
$U(1)$ variables. This happens again because the $Z_6$ symmetry binds
$U(1)$ variables with the center of the $SU(3)$ subgroup of the gauge
group.

As in \cite{BVZ2004} we mention here that the $SU(2)$ fundamental Higgs
model, has a similar phase structure as our model, except for the
absence of the phase transition line between phases I and II. In the
latter model it was shown that different phases are actually not
different. This means that the phase transition line ends at some point
and the transition between two states of the model becomes continuous.
Thus one may expect that in our model the phase transition line between
phases I and III ends at some point. However, we do not observe this
for the considered values of couplings.

In our model both phase transition lines join in a triple point,
forming a common line. This is, evidently, the consequence of the
mentioned additional symmetry that relates $SU(2)$ , $U(1)$, and
$SU(3)$  excitations. The same picture, of course, does not emerge in
the conventional $SU(3) \otimes SU(2) \otimes U(1)$ gauge -- Higgs
model: its $SU(2) \otimes U(1)$ part was investigated, for example, in
\cite{SU2U1}. As for the $SU(3)$ gauge theory, it has no phase
transition at finite $\beta$ and zero temperature at all.

\section{Conclusions}
\label{sect.7}

We summarize our results as follows:

\begin{enumerate}
\item We performed a numerical investigation of the quenched lattice
model that respects the additional symmetry.

\item The lattice model contains three phases. In the first phase the
potential between static leptons is confinement-like.  In the second
phase the confining forces are observed, at sufficiently small
distances, between the left-handed external leptons.  The last one is
the Higgs phase, where there are no confinement - like forces between
static leptons at all.

\item Investigation of the monopoles constructed of colored fields shows
that colored fields feel the phase transition lines.

\item In all phases of the model we observe confinement of quarks. The
string tensions for different kinds of quarks are equal to each other.

\item The main consequence of the emergence of the additional symmetry
is that the phase transition lines corresponding to the $SU(2)$ and
$U(1)$ degrees of freedom join in a triple point forming a common line.
This reflects the fact that the $SU(2)$ and $U(1)$ excitations are
related due to the mentioned symmetry. The same situation does not
occur in the conventional $SU(2)\otimes U(1)$ gauge - Higgs model
\cite{SU2U1}.
\end{enumerate}

So, we have found a qualitative difference between the conventional
discretization and the discretization that respects the invariance
under the transformations given in Eq.~(\ref{sym}).

In order to illustrate other possible differences let us consider the
problem of constructing the operator which creates a glueball in the
$Z_6$ invariant version of the lattice SM. Here we cannot use the
conventional expression
\begin{equation}
 {\cal O}_c =
 \left(1 - \frac{1}{3}{\rm Re}\, {\rm tr}\, \Gamma_{\rm plaq}\right)
\label{GCO0}
\end{equation}
as it is not invariant under our $Z_6$ symmetry. Instead we may use
$Z_6$ - invariant expressions like
\begin{equation}
 {\cal O} = \left(1- \frac{1}{3}{\rm Re}\, {\rm tr} \{\Gamma_{\rm plaq}
 e^{-\frac{2i}{3}\theta_{\rm plaq}}\}\right) - \frac{1}{9}( 1 -{\rm cos}\, 2
 \theta_{\rm plaq})\label{GCO1}
\end{equation}
In the naive continuum limit the above expressions (\ref{GCO0}) and
(\ref{GCO1}) coincide. In a similar way the naive continuum limit of
the action (\ref{Act}) coincides with that of the conventional lattice
SM action for the appropriate choice of coupling constants.

However, this coincidence does not mean necessarily, that either the
models themselves or the correlators of operators (\ref{GCO0}) and
(\ref{GCO1}) lead to the same results. Let us recall here two
precedents, i.e., two similar situations, where the coincidence of the
naive continuum limits does not lead to the same physics.

The first example is the massless lattice fermion. One may compare Wilson
fermions with the simplest direct discretization of the Dirac fermion
action\cite{lattice_fermions}. These two actions differ from each other
by a term which naively vanishes in the continuum limit. However,
the corresponding models are not identical from the physical point of
view. Namely, the second one contains $15$ additional fermion species
while in the Wilson formulation all of them acquire infinite mass and
disappear in the continuum limit. This phenomenon of fermion
doubling is widely discussed in the literature.  It is worth
mentioning that another difference between these two formulations is
the absence of exact chiral symmetry in the Wilson formulation and its
appearance in the naive discretization.

The second example is the pure nonabelian gauge theory. If we would
discretize its form written in terms of gauge potentials losing the
exact  gauge invariance, the resulting lattice model would have the
same naive continuum limit as the conventional lattice gluodynamics,
which is written in terms of link matrices. However, in such a
definition of lattice gauge theory confinement is
lost\cite{noncompact}.

In the two examples of lattice models considered above, which have the
same naive continuum limit but different symmetry properties, finally
lead to different physics. Exactly the same situation may be present in
our case, where the naive continuum limit of the two lattice
realizations of the SM is the same, while only one formulation is $Z_6$
invariant.

Another argument in favor of the point of view that these two models
are indeed different, comes from the direct consideration of how
continuum physics emerges in the lattice SM. Namely, there are
indications \cite{BVZ2002,GZP} that several kinds of singular field
configurations may survive in the continuum limit of nonabelian lattice
gauge models. If so, the conventional action of the lattice SM and the
action (\ref{Act}) may appear to be different when approaching the
continuum for singular field configurations of various kinds.

\vspace{5ex}

We are grateful to F.V. Gubarev and Yu.A. Simonov for useful
discussions. A.I.V.  and M.A.Z. kindly acknowledge the hospitality of
the Department of Physics and Astronomy of the Vrije Universiteit,
where part of this work was done.  We also appreciate I.~Gogoladze and
R.~Shrock, who have brought to our attention references \cite{Z6} and
\cite{SU2U1} respectively. This work was partly supported by RFBR
grants 03-02-16941, 05-02-16306,  04-02-16079, and PRF grant for
leading scientific schools N 1774.2003.2, by Federal Program of the
Russian Ministry of Industry, Science, and Technology No
40.052.1.1.1112.


\begin{thebibliography}{99}

\bibitem{BVZ2003}
B.L.G.~Bakker, A.I.~Veselov, and M.A.~Zubkov, Phys. Lett. B {\bf  583}, 379
(2004);

\bibitem{BVZ2004}
B.L.G.~Bakker, A.I.~Veselov, and M.A.~Zubkov, Yad. Fiz.  {\bf  }, (2005);

\bibitem{Z6}
K.S.~Babu, I.~Gogoladze, and K.~Wang, Phys. Lett. B {\bf 570}, 32 (2003);\\
K.S.~Babu, I.~Gogoladze, and K.~Wang, Nucl. Phys. B {\bf 660}, 322 (2003);

\bibitem{SU2U1}
R.~Shrock, Phys. Lett. B {\bf 162}, 165 (1985); Nucl. Phys. B {\bf 267}, 301
(1986).

\bibitem{Greensite}
J.~Greensite, Prog. Part. Nucl. Phys. {\bf 51} (2003) 1

\bibitem{BVZ1999}
B.L.G.~Bakker, A.I.~Veselov, and M.A.~Zubkov, Phys. Lett. B {\bf 471} (1999) 214

\bibitem{forms}
M.I. Polikarpov, U.J. Wiese, and M.A. Zubkov, Phys. Lett. B {\bf 309}, 133
(1993).

\bibitem{BVZ2002}
B.L.G.~Bakker, A.I.~Veselov, and M.A.~Zubkov, Phys. Lett. B {\bf 544}, 374
(2002);

\bibitem{GZP}
F.V.~Gubarev and V.I.~Zakharov, hep-lat/0211033;\\
F.V.~Gubarev, A.V.~Kovalenko, M.I.~Polikarpov, S.N.~Syritsyn, and V.I.~Zakharov,
Phys. Lett. B {\bf 574} 136 (2003)

\bibitem{Montvay}
I.~Montvay, Nucl. Phys. {\bf B269}, 170 (1986).

\bibitem{lattice_fermions}
N.B.~Nielsen and M.~Ninomiya, 
Nucl. Phys. B {\bf 185}, 20 (1981); {\it ibid}, 173;\\
M.~L{\"{u}}scher, Phys. Lett. B {\bf 428}, 342 (1998);\\
H.~Neuberger, Phys. Lett. B {\bf 417}, 141 (1998)\\

\bibitem{noncompact}
A.~Patrascioiu, E.~Seiler, and I.O.~Stamatescu, 
Phys. Lett. B {\bf 107}, 364 (1981); \\
E.~Seiler, I.O.~Stamatescu, and D.~Zwanziger, 
Nucl. Phys. B {\bf 239}, 177 (1984); \\
Y.~Yotsuyanagi, Phys. Lett. B {\bf 135}, 141 (1984); \\
K.~Cahill, S.~Prasad, R.~Reeder, and B.~Richter,  \\
Phys. Lett. B {\bf 181}, 333 (1986); \\
K.~Cahill, Phys. Lett. B {\bf 231}, 294 (1989)

\bibitem{EW_T}
M.~Gurtler, E.M.~Ilgenfritz, and A.~Schiller, Phys. Rev. D {\bf 56}, 3888
(1997);\\
B.~Bunk, E.M.~Ilgenfritz, J.~Kripfganz, and A.~Schiller, Nucl. Phys. B {\bf
403}, 453 (1993);\\
M.N.~Chernodub, F.V.~Gubarev, E.M.~Ilgenfritz, and A.~Schiller, Phys. Lett. B
{\bf 434}, 83 (1998);\\
M.N. Chernodub,~F.V. Gubarev, E.M.~Ilgenfritz, and A.~Schiller, Phys. Lett. B
{\bf 443}, 244 (1998).\\

\end{thebibliography}
\end{document}